\documentclass{appolb}
\usepackage{graphicx}

\begin{document}
\title{Chiral fluctuations in MnSi above the Curie temperature measured 
with polarized inelastic neutron scattering
\thanks{Presented at the Strongly Correlated Electron Systems 
Conference, Krak\'ow 2002}%
}


\author{B.~Roessli
\address{Laboratory for Neutron Scattering, ETH Zurich \& Paul
Scherrer Institute, CH-5232 Villigen PSI}
\and
P.~B\"oni, W.~E.~Fischer, and Y.~Endoh
\address{ Physik-Department
E21, Technische Universit\"at M\"unchen, D-85747 Garching, Germany
\\ Laboratory for Neutron Scattering, ETH Zurich \& Paul
Scherrer Institute, CH-5232 Villigen PSI\\Institute of Materials Research, Tohoku University, Katahira, Aoba-ku,
Sendai, 980-8577, Japan}
}
\maketitle


\begin{abstract}
Using polarized inelastic neutron scattering  the antisymmetric part of
the dynamical susceptibility in non-centrosymmetric MnSi is determined. The
paramagnetic fluctuations are found to be incommensurate with the
chemical lattice and to have a chiral character. We show that   
antisymmetric interactions must be taken into account to properly
describe the critical dynamics in MnSi above the Curie temperature. 
The inelastic neutron data is interpreted within the framework of the 
SCR-theory, taking into account the Dzyaloshinskii-Moriya interaction. 
\end{abstract}


  
\section{Introduction}
Ordered states with helical arrangement of the magnetic moments
are described by a chiral order parameter $\vec C=\vec S_1 \times \vec S_2$,
which yields the left- or right-handed rotation of neighboring spins along
the pitch of the helix. The detection of chiral fluctuations is, however, a
difficult task and it is only recently
that chiral fluctuations could be observed in triangular antiferromagnets
with polarized-neutron scattering when
an external magnetic field is applied \cite{maleyev95,plakhty99}.
\\
The metallic compound MnSi crystallizes in the cubic space group P2$_1$3
that lacks
a center of symmetry.  The Curie temperature is $T_C =
29.5$ K. 
Below $T_C$  the magnetic moments build a ferromagnetic spiral along the
[1 1 1] direction with a period of approximately 180 \AA. 
The spontaneous magnetic moment of Mn is $\mu
\simeq 0.4 \mu_B$ that is strongly reduced from the free ion value
of $2.5\mu_B$. 
The four Mn atoms are placed at
the positions $(u,u,u)$, $({1\over2}+u,{1\over2}-u,-u)$, 
$({1\over2}-u,{1\over2}+u,-u)$,
$({1\over2}+u,-u,{1\over2}-u)$.
Being a prototype of a weak itinerant ferromagnet, the magnetic
fluctuations in MnSi have been investigated in detail by means of
polarized \cite{tixier} and unpolarized neutron scattering
\cite{ishikawa_85}.

\section{Experimental}
We investigated the paramagnetic fluctuations in a single crystal of MnSi
 on the triple-axis spectrometer TASP at the neutron
spallation
source SINQ. The spectrometer was operated in the constant final
energy mode with a  neutron wave vector $\vec k_f$=1.97 $\AA^{-1}$. In
order to suppress contamination by higher order neutrons a
pyrolytic-graphite filter was
installed in the
scattered beam. 
To polarize the incident neutron beam a remanent FeCoV/TiN-type
bender was inserted after the monochromator. 
The advantage of such a device is that to reverse the 
spin of the neutron no spin flipper is necessary thanks to the remanent magnetization of
the super-mirror coatings of the benders~\cite{boni}.
The polarization of
the neutron
beam was maintained along the neutron path by a guide field B$_g$=10G that
defines the polarization of the neutrons $\vec P_0$ with respect to the   
scattering vector $\vec Q$. 
We did not analyze the polarization of
the scattered neutrons during the course of these experiments.
The sample was mounted in a standard $\rm ^4He$ orange-cryostat of ILL-type 
with the [1 0 0] and [0 1 1] crystallographic in the scattering plane. 
The measurements were performed around the (0 1 1) Bragg reflection in the 
paramgagnetic phase. 
 
\section{Results}
A typical constant-energy scan at $\hbar \omega$=0.5 meV 
and T=35K using a polarized beam as described above  
is shown in Fig.\ref{Fig1}.
In a first step we chose the polarization of the neutron beam
along the scattering vector $\vec Q$, repeated the measurements with the
polarization aligned along $-\vec Q$ and then calculated the difference 
between the two sets of measurements. 
It is obvious from Fig.~\ref{Fig1} that the inelastic scattering is
polarization dependent. Of particular importance, we find that 
the neutron peaks appear at positions incommensurate  
with respect to the chemical lattice, namely at
 $\vec Q=\vec \tau\pm \vec\delta$ ($\vec\tau$ is
a reciprocal lattice vector). 

In order to discuss our results we start with the general
expression for the cross-section of magnetic scattering with
polarized neutrons
\begin{equation}  
{d^2\sigma\over{d\Omega d\omega}} \sim  \sum_{\alpha,
\beta}(\delta_{\alpha,\beta}
\hat Q_\alpha \hat Q_\beta) A^{\alpha \beta} (\vec Q, \omega)
+  \sum_{\alpha, \beta} (\hat {\vec Q} \cdot \vec
P_i)\sum_{\gamma}\epsilon_{\alpha, \beta, \gamma} \hat Q^\gamma
B^{\alpha \beta}(\vec Q, \omega) \label{ncs}
\end{equation}
where $(\vec Q, \omega)$ are the momentum and energy-transfers
from the neutron to the sample, $\hat {\vec Q} = \vec Q/|\hat Q|$,
and $\alpha, \beta, \gamma$ indicate Cartesian coordinates. The   
first term in Eq.~\ref{ncs} is independent of the polarization of
the incident neutrons, while the second is polarization dependent 
through the factor $(\hat{\vec Q} \cdot \vec P_i)$. $\vec P_i$   
denotes the direction of the neutron polarization and its scalar  
is equal to 1 when the beam is fully polarized.
It can be shown~\cite{lovesey_99} that $A^{\alpha \beta}$
and $B^{\alpha \beta}$ are the symmetric and antisymmetric part of
the scattering function $S^{\alpha \beta}$, that is $A^{\alpha
\beta}={1\over 2} (S^{\alpha \beta} + S^{\alpha \beta})$ and
$B^{\alpha \beta}={1\over 2} (S^{\alpha \beta} - S^{\beta
\alpha})$. $S^{\alpha \beta}$ are the Fourier transforms of the   
spin correlation function $<s^\alpha_l s^\beta_{l'}>$, $S^{\alpha
\beta}(\vec Q, \omega)={1\over{2\pi N}}\int_{-\infty}^\infty{dt
e^{-i\omega t} \sum_{ll'}{e^{i\vec Q (\vec X_l-\vec X_{l'})
}}<s^\alpha_l s^\beta_{l'}(t)> }$. The vectors $\vec X_l$ designate 
the positions of the scattering centers in the lattice.
Accordingly,  a finite contribution to the antisymmetric part of the
wave-vector dependent dynamical susceptibility has been measured in 
MnSi in the paramagnetic phase.

\section{Discussion and Conclusion}
\begin{figure}[!ht]
\begin{center}
\includegraphics[width=0.6\textwidth, angle=-90]{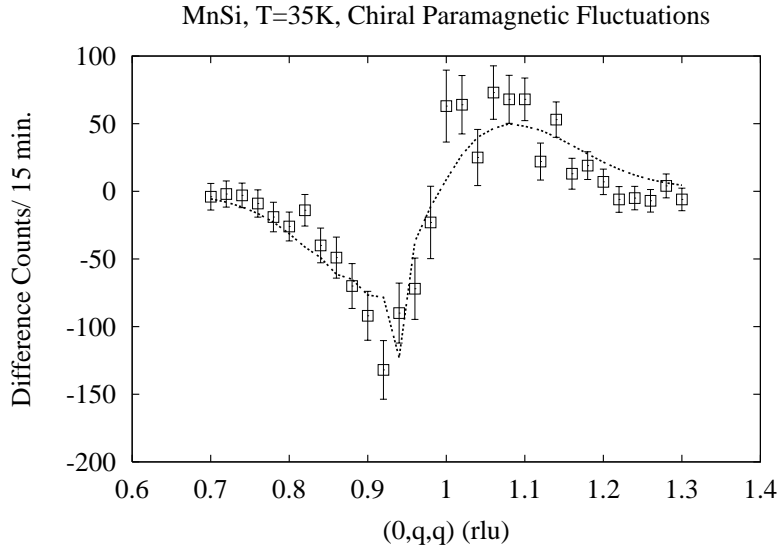}
\end{center}
\caption{Polarization dependent part of the dynamical
susceptibility measured in MnSi at T=35K and $\hbar \omega$=0.5meV. The line is a fit to the data.}
\label{Fig1}
\end{figure}

In a similar way as  for insulators with
localized spin densities, we interpret 
the antisymmetric interaction observed in MnSi as originating
from the spin-orbit coupling in the absence of an inversion center. 
In that case, the polarisation dependent part of the neutron cross-section 
is given by  
\begin{equation} 
 \biggl({{d^2\sigma}\over{d\Omega d\omega}}\biggr)_{\Delta}
   \sim (\vec D \cdot \hat{\vec Q})(\hat{\vec Q}\cdot \vec P_i) 
   \Im{{(\chi^\perp(\vec q-\vec\delta,\omega)-\chi^\perp (\vec q+\vec\delta,\omega))}\over
   {2D}}.
\label{pa}
\end{equation}
The transverse
susceptibilities in Eq.~\ref{pa}  for itinerant
magnets as given by self-consistent re-normalization theory (SCR) are 
\cite{moriya_85}
\begin{equation}
 \chi^\perp (\vec q \mp \vec \delta, \omega) =
           \chi^\perp (\vec q \mp \vec \delta)/(1-i\omega/\Gamma_{\vec q \mp \vec \delta}).
\label{src}
\end{equation}
$\vec \delta$ is the ordering wave vector, $\chi^\perp (\vec q \mp
\vec \delta)=\chi^\perp(\vec \pm \delta)/(1+q^2/\kappa^2_\delta)$ 
the static susceptibility, and $\kappa_\delta$ the inverse
correlation length. For itinerant ferromagnets the damping of the
spin fluctuations is given by $ \Gamma_{\vec q \mp \vec \delta} =
uq (q^2+\kappa^2_\delta)$ with $u = u(\vec \delta)$ reflecting the 
damping of the spin fluctuations.
Here, $\vec q$ designates the reduced momentum transfer with
respect to the nearest magnetic Bragg peak at $\vec \tau \pm \vec
\delta$. The solid line of Fig.~\ref{Fig1} shows calculations of                    
$({{d^2\sigma}/({d\Omega d\omega}}))_{\Delta}$ as defined above which describes 
the inelastic polarised neutron data well with  $\kappa_0$ = 0.12$ \AA^{-1}$
and $u = 27$  
meV$\AA^3$~\cite{roessli_02}. In conclusion, we have shown that 
in MnSi the axial interaction leading to 
the polarized part of the neutron cross-section can  
been identified as originating from the DM-interaction.

\end{document}